\documentclass[rmp,10pt,byrevtex,notitlepage,secnumarabic,nofootinbib,a4paper,longbibliography,showkeys]{revtex4-1}
\usepackage{graphicx}
\usepackage{amsmath}
\usepackage{amssymb}
\usepackage{float}
\usepackage{url}
\usepackage{xr}
\usepackage{setspace}

\usepackage{xcolor}
\usepackage{colortbl}
\usepackage[colorlinks=true,urlcolor=blue,linkcolor=blue,citecolor=blue]{hyperref}
\usepackage{siunitx}
\usepackage{mathtools}
\usepackage[utf8]{inputenc}
\usepackage[english]{babel}
\usepackage{array}
\AtBeginDocument{\usepackage{booktabs}}
\graphicspath{{./figures/}{./figures/relojes/}}

\definecolor{BLUE2}{HTML}{084594}
\begin{document}
\author{José María Martín-Olalla}
\address{Universidad de Sevilla. Facultad de Física. Departamento de Física de la Materia Condensada. Sevilla, Spain}
\email{email: olalla@us.es. Twitter: @MartinOlalla\_JM}
\date{Jul 29th, 2019}
\title{Scandinavian bed and rise times in the Age of Enlightenment and in the 21st century show similarity, helped by Daylight Saving Time}
\begin{abstract}
  This is a pre-print of the Letter to the Editor to be published in the \emph{Journal of Sleep Research} (2019) e12916 doi: \href{http://dx.doi.org/10.1111/jsr.12916}{10.1111/jsr.12916}. 
\end{abstract}

\maketitle

The past few years have witnessed a growing interest in the  impact of artificial light on human sleep timing and human sleep ecology. Scores of reports have characterized sleep timing in pre-industrial societies\citep{Knutson2014,Yetish2015,Moreno2015,DeLaIglesia2015,Beale2017,Samson2017,Pilz2018} and in industrial societies\citep{Monsivais2017a,Martin-Olalla2019b}. In some of these reports the role of latitude and seasonal variations is analyzed.

Recently the \emph{Journal of Sleep Research} has presented\citep{VanEgmond2019} the bed and rise times of Olof Hiorter (1696-1750), a Swedish astronomer based on Uppsala, at \ang{60} latitude, which continued the work by Anders Celsius on Earth's magnetic field. It is a most interesting piece because data cover one calendar year; latitude and seasonal variations of light and dark are extreme; and speaks about human behaviour in pre-modern times, where the impact of articial light, social timing and alarm clocks is expected to be less much less significant.

Authors found a natural seasonal spread of bed and rise times and noted that  ``seasonal effects on sleep are absent in our modern electrical lighting environment, where the contrast between daytime and evening light has become less pronounced''. Yet, and as far as bed and rise times are concerned, this is a misunderstanding: industrial, urban societies have pushed for regular, year round social timing but, frequently, within an environment where Daylight Saving Time regulations occur.  If anything else changes, DST provides seasonal variability to rise and bed times. Time in Sweden follows DST regulations since the 1980s. 

Present day rise and bed times in Sweden can be extracted from the Harmonized European Time Use Survey (hereafter \textsc{Hetus}), a collection of Time Use Surveys in fifteen European countries compiled and harmonized by, incidentally Statistics Sweden and Statistics Finland. The collection includes data from the Baltic States, Finland and Norway.  \textsc{Hetus} pre-preparred tables report the daily rhythm of \emph{sleep and other personal care}: the percentage of respondents sleeping or doing personal care, like dressing, undressing or taking a shower, as a function of time in one day. The daily rhythm looks like a window function, soaring at night, and plummeting at dawn. From these transitions rise times and bed times can be evaluated as described by \citet{Martin-Olalla2018}.

On the other hand, Hiorter's bed and rise times were reported on a monthly basis\cite{Ekman2018}. These values can be grouped in two average values: winter (September to February) and summer (March to August). The year is partitioned in this specific way because months are referred to the Julian calendar, official by that time in Sweden, with equinoctes coming eleven days earlier in the calendar date.

\begin{figure*}[t]
  \centering
  \includegraphics[width=\textwidth]{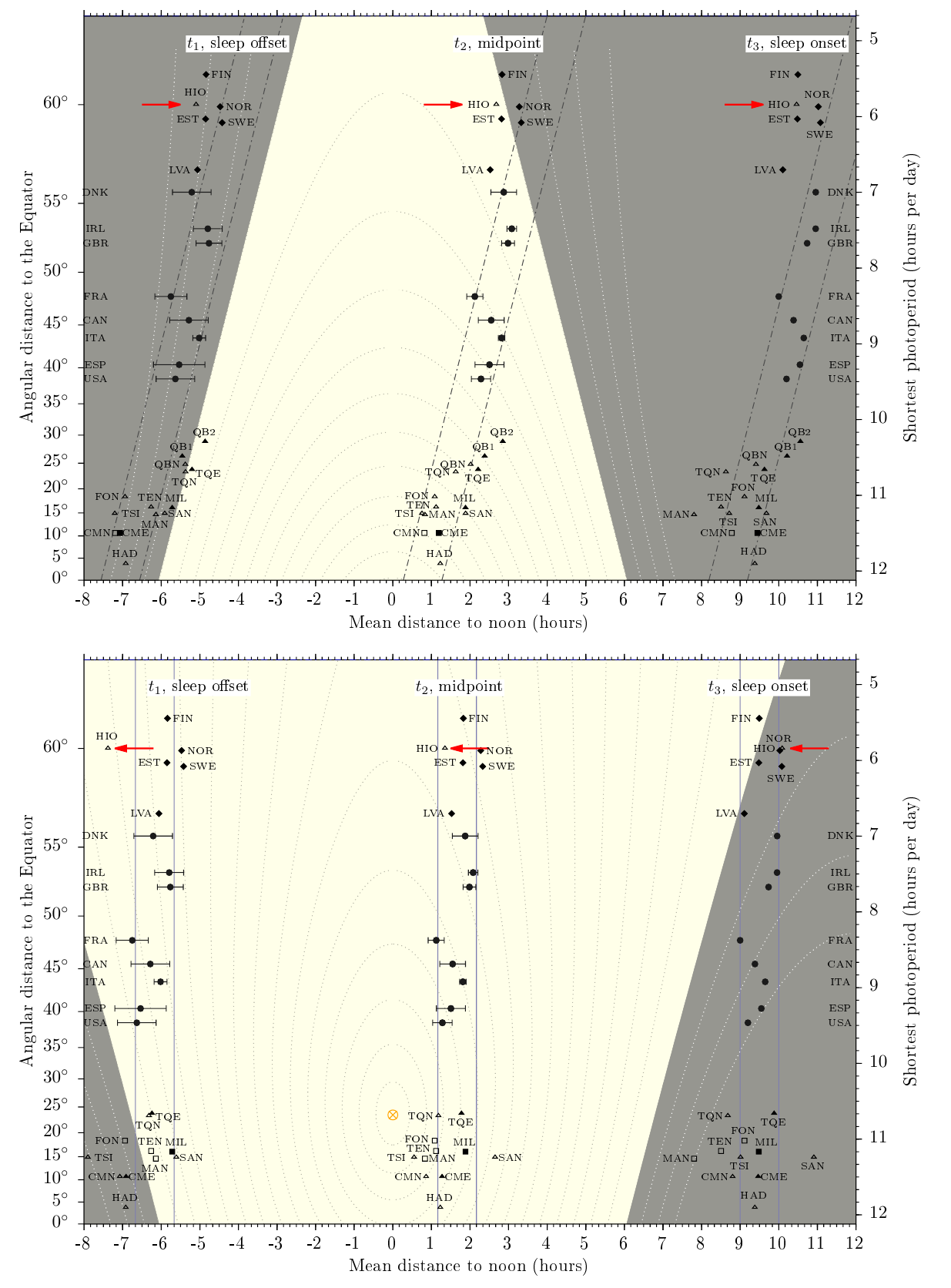}
  \caption{Human sleep ecology as a function of latitude and the light  and dark conditions. Top panel shows winter setting; bottom panel, summer setting. \protect\citet{Martin-Olalla2019b} gives a detailed description of the panels. Diamonds display \textsc{Hetus} values. Hiorter's values (see~\protect\citet{VanEgmond2019} and \protect\citet{Ekman2018}) are labeled as \textsc{hio} and highlighted by red arrows. Values from industrial, urban societies (circles and diamonds) take into account Daylight Saving Time in the bottom panel.}
  \label{fig:sleep}
\end{figure*}

Figure~\ref{fig:sleep} shows Scandinavian and Baltic data from \textsc{Hetus} (diamonds) and from Hiorter’s bed and rise times (noted by arrows), altogether with previously collected data from pre-industrial (triangles and squares) societies and industrial (circles) societes (see \citet[Fig~2]{Martin-Olalla2019b}). In the figure bold symbols highlight use of efficient, artificial light; whereas open symbols highlight lack of artificial light. The top panel shows the winter season; the bottom panel, summer.

The only significant difference between Hiorter’s rise and bed times, and present day times in Sweden occur for summer rise times.

Daylight saving time is being heavily criticized by physiologists and medical doctors due to the annoying, biannual transitions that it involves and their impact on the circadian system and public health (see \citet{Watson2019,Roenneberg2019} and \citet{MeiraeCruz2019}). Indeed, modern, urban societies have lost their ability to smoothly track sunrises, a characteristic in Hiorter's behaviour. That is a consequence of social timing having been synced to clock time. Upon this scenario, the shift of one hour seems the only practical method to achieve the natural advance in the phase of human sleep timing, which is present in this case report from the Age of Enlightenment. That helps understanding why DST succeeded in industrial, urban societies.

\acknowledgements

The author expresses his gratitude to Statistics Sweden and Statistics Finland for giving access to \textsc{Hetus} data. They are available at \href{https://www.h6.scb.se/tus/tus/}{https://www.h6.scb.se/tus/tus/}.

This project was started on Jul 22nd, 2019 after a \href{https://www.researchgate.net/publication/332575275_Bed_and_rise_times_during_the_Age_of_Enlightenment_A_case_report}{suggestion} appeared in author's homepage at ResearchGate social network. This work was not founded by any means. No conflict of interest exists.

\bibliographystyle{plainnat}

\end{document}